\documentclass[twocolumn,showpacs,superscriptaddress,amsmath,amssymb,prl]{revtex4}

\usepackage{graphicx}%
\usepackage{dcolumn}%

\begin{document}

\title{Details of the photoemission spectra analysis}

\author{A. A. Kordyuk}
\affiliation{Institute for Solid State Research, IFW-Dresden, Helmholtzstr.~20, D-01069 Dresden, Germany}
\affiliation{Institute of Metal Physics of National Academy of Sciences of Ukraine, 03142 Kyiv, Ukraine}

\author{S. V. Borisenko}
\affiliation{Institute for Solid State Research, IFW-Dresden, Helmholtzstr.~20, D-01069 Dresden, Germany}

\date{September 30, 2005}%

\begin{abstract}
Here we present some details of the self-consistent procedure of the photoemission spectra analysis suggested in [Phys.~Rev.~B \textbf{71}, 214513 (2005); cond-mat/0405696; cond-mat/0409483] and answer some of the most frequently asked questions concerning this analysis.
\end{abstract}

\pacs{74.25.Jb, 74.72.Hs, 79.60.-i, 71.18.+y}%

\maketitle

\subsection{1. Quadratic dispersion}

\textit{How one can derive Eq.~(5) in Ref.\onlinecite{KordyukPRB2005}?}

We start from the spectral function
\begin{eqnarray}\label{A}
A(\omega, k) = -\frac{1}{\pi}\frac{\Sigma''(\omega)}{(\omega - \varepsilon(k) - \Sigma'(\omega))^2 + \Sigma''(\omega)^2},
\end{eqnarray}
where $\varepsilon(k)$ is the bare band dispersion along a certain direction in the reciprocal space. Here $\Sigma''(\omega) < 0$, and $\Sigma'(\omega) > 0$ for $\omega < 0$.

\begin{figure}[b]
\includegraphics[width=8.47cm]{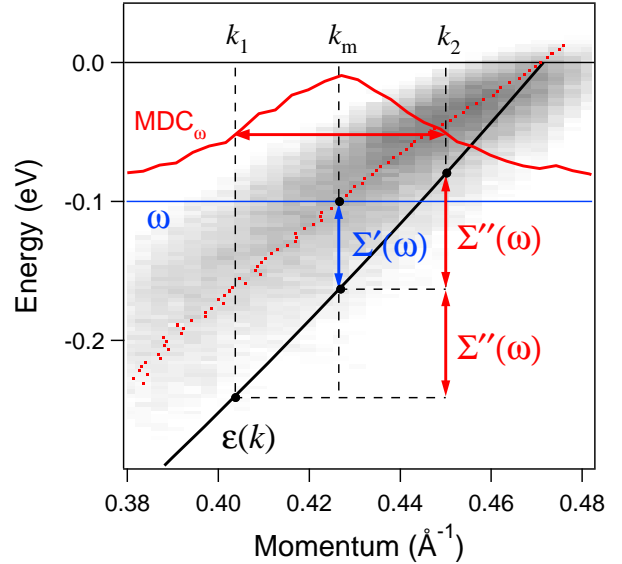}%
\caption{\label{Fig1} Real and imaginary parts of the self-energy (shown by blue and red double headed arrows respectively) on a photoemission image. Bare band dispersion (black solid line) and renormalized dispersion (red points). Red solid line represents the momentum distribution curve (MDC) taken at $\omega$. The position of MDC  maximum determines $k_m$, and the momenta of its half maximum level determine $k_1$ and $k_2$.}
\end{figure}

For each fixed energy $\omega$ we define three momenta: $k_m(\omega)$, $k_1(\omega)$, and $k_2(\omega)$ as 
\begin{eqnarray}
A(k_m) &=& \max[A(k)],\\
A(k_{1,2}) &=& \max[A(k)]/2. 
\end{eqnarray}
Taking $k_2 > k_1$, we define the MDC width $2W = k_2 - k_1$. Solving these equations one gets
\begin{eqnarray}
\label{km}\omega - \varepsilon(k_m) - \Sigma' &=& 0,\\
\label{k12}\omega - \varepsilon(k_{1,2}) - \Sigma' &=& \pm|\Sigma''|,
\end{eqnarray}
from where three expressions for $\Sigma''$ can be derived
\begin{eqnarray}\label{ImS}
|\Sigma''| &=& \varepsilon(k_m) - \varepsilon(k_1)\\
&=& \varepsilon(k_2) - \varepsilon(k_m)\\
&=& [\varepsilon(k_2) - \varepsilon(k_1)]/2.
\end{eqnarray}
Fig.~\ref{Fig1} illustrates this. These expressions are general and exact. One can use any of them, depending on what one determines in the experiment. We prefer to deal with two experimental quantities, the renormalized dispersion $k_m(\omega)$ and MDC width $2W(\omega)$. So, we focus on the last equation. 

For a linear bare dispersion, $\varepsilon(k) = v_F (k - k_F)$, this gives simple
\begin{eqnarray}\label{ImS_L}
|\Sigma''| = v_F W.
\end{eqnarray}

For a quadratic dispersion, $\varepsilon(k) = \omega_0 [(k/k_F)^2 - 1]$,
\begin{eqnarray}\label{ImS_Q}
|\Sigma''| &=& \frac{\omega_0}{2 k_F^2}(k_2^2-k_1^2)\nonumber\\
&=& \frac{\omega_0}{k_F^2} W (k_2+k_1)\nonumber\\
&=& \frac{\omega_0}{k_F^2} 2W \sqrt{k_m^2-W^2},
\end{eqnarray}
here $\omega_0 >0$. In the last step we used the property of a parabolic function $\varepsilon(k)$:
\begin{eqnarray}\label{para}
\text{if} & 2\varepsilon(k_m) &= \varepsilon(k_2) + \varepsilon(k_1),\nonumber\\
\text{then} & 4 k_m^2 &= (k_2+k_1)^2 + (k_2-k_1)^2.
\end{eqnarray}

It is interesting here to introduce an asymmetry factor $L_a$ which can be simply determined from experiment but keeps an information about underlying bare dispersion:
\begin{eqnarray}\label{asym}
L_a = \frac{\sqrt{k_m^2-{\langle k \rangle}^2}}{W},
\end{eqnarray}
where ${\langle k \rangle} = (k_2+k_1)/2$ and  $W = (k_2-k_1)/2$. For parabolic dispersion $L_a = 1$.

\subsection{2. Renormalized velocity}

\textit{Is it physical for the renormalized velocity to be infinite?} 

From a formal point of view, one cannot even write an expression for the renormalized dispersion
$\varepsilon_m(k)$, but only for its inverse function $k_m(\omega)$. E.g., for a linear bare dispersion, $\varepsilon = v_F (k - k_F)$, from Eq.~(\ref{km}), 
\begin{eqnarray}\label{depsilon}
v_F \frac{dk_m}{d\omega} = 1 - \frac{d\Sigma'}{d\omega}.
\end{eqnarray}
So, $d\varepsilon_m(k)/dk = \infty$ just when $d\Sigma'/d\omega = 1$, i.e.~when $|\Sigma'(\omega)|$ decreases faster than $\omega$ ($\Sigma'(\omega) > 0$ for $\omega < 0$). Such a situation is quite natural because any sharp step in $\Sigma''(\omega)$ results in a discontinuity in $\Sigma'(\omega)$.

\subsection{3. Scattering on impurities}

\textit{What is the reason for the offset of the scattering rate?}

A general answer is that the scattering rate determined from photoemission experiment consists of a finite experimental resolution $Res$, non-zero temperature contribution of the self-energy  $\Sigma''(0,T)$, and a scattering on impurities $\Sigma''_{imp}$, but it is rather difficult to disentangle all of them. The experimental resolution is the main reason for this, since it depends not only on resolution of the analyser and a light source, $Res_0$ but also on quality of sample surface, $Res_s$, and highly varies from sample to sample. Therefore, at present stage, we can only believe that accumulating the photoemission spectra from a number of samples and cleaves we approach the $Res \rightarrow Res_0$ limit for the sharpest spectra. And it is clear that for these spectra the scattering on impurities, if sample-dependent, is also the lowest one. This means that in such a simple consideration we can only estimate a maximum limit for a minimal impurity scattering.

The minimal value for the full width at half a maximum (FWHM) of $E_F$-MDC at the nodal point is $2W \approx$ 0.013 \AA$^{-1}$ \cite{KordyukPRL2004}. Using for an estimate Eq.~(\ref{ImS_L}) and $v_F =$ 4 eV\AA~gives an effective $\Sigma''_{eff}(0) \approx$ 26 meV. 

The momentum resolution, derived from the angular resolution of the analyser about 0.1$^\circ$, $R_k \approx$ 0.007 \AA$^{-1}$ or 14 meV (here the renormalized velocity $v_R \approx 2$ eV\AA~should be used as a coefficient). The energy resolution $R_\omega \approx$ 12 meV. Then, in case $R_k$ and $R_\omega$ are comparable, the overall resolution can be estimated as $Res = \sqrt{R_{\omega}^2+R_k^2} \approx$ 18 meV.

The contribution of temperature into self-energy is most likely associated with the primary scattering channel due to direct electron-electron interactions (Auger-like decay) \cite{KordyukPRL2004}. For this channel $-\Sigma''(\omega, T) = \alpha [\omega^2 + (\pi T)^2]$, and taking $\alpha \approx 2$, $-\Sigma''(0) \approx$ 13 meV at 300 K, but is negligible $\approx$ 0.13 meV at 30 K.

Thus, the above mentioned limit of the self-energy due to scattering on impurities at low temperature can be evaluated as
\begin{equation}
\label{Impurity}\Sigma''_{imp} = \Sigma''_{eff}(0) - Res \approx 8 \textnormal{ meV}.
\end{equation}

It is interesting that with time due to surface aging $\Sigma''_{eff}(0)$ become essentially higher, that can be explained by growing $\Sigma''_{imp}$ 2-3 times of its initial value. Therefore, the increase of the impurity scattering with aging as well as with temperature can be in principle investigated.

Another possibility comes from lineshape analysis. We have noticed that when approaching the Fermi level the MDC lineshape demonstrates a crossover from lorentzian to gaussian, such as the sharpest $E_F$-MDCs better fit to an approximately equal mixture of both. This not only validates the above estimation but also shows a way of a careful evaluation of the constituents of $\Sigma''_{eff}(0)$, which, although, requires a very high experimental statistics.

\subsection{4. Resolution effect}

\textit{How the resolution is taken into account?}

In \cite{KordyukPRB2005} we consider a complex $\omega$-dependent contribution of the resolution, $R(\omega) = R'(\omega) + iR''(\omega)$, to $\Sigma_{width}(\omega)$, the self-energy determined from the MDC width. We define $\Sigma''_{width}(\omega)$ through some real $\Sigma(\omega)$ as $\Sigma''_{width}(\omega) = \sqrt{\Sigma''(\omega)^2 + R^2}$, where $R$ is an overall resolution parameter. Subsequently,
\begin{eqnarray}
\label{ImR} R''(\omega) &=& \sqrt{R^2 + \Sigma''(\omega)^2} - \Sigma''(\omega),
\end{eqnarray}
and, through the Kramers-Kronig (KK) transforamation,
\begin{eqnarray}
\label{ReR} R'(\omega) &=& \mathbf{KK} R''(\omega).
\end{eqnarray}

In \cite{KordyukPRB2005} we derive $R'(\omega)$ empirically using $R$ as a parameter and consider only the variable part of $\Sigma''$ in Eq.~\ref{ImS}, i.e.~without the offset: $\Sigma''(\omega)-\Sigma''(0)$. With such a procedure $R$ = 0.015 eV.

In principle, the resolution effect can be explicitly calculated from the known energy and momentum resolutions. If one knows the overall resolution, e.g.~$Res = \sqrt{R_{\omega}^2+R_k^2}$, where $R_{\omega}$ and $R_k$ are the energy and momentum resolutions (in energy units) respectively, then
\begin{eqnarray}\label{GoodRes}
R''(\omega) = \Sigma''_{width}(\omega) - \sqrt{\Sigma''_{width}(\omega)^2-Res^2}.
\end{eqnarray}

\section{Erratum}

We correct two missprints in Ref.~\onlinecite{KordyukPRB2005}: should be $-v_F$ instead of $v_F$ in Eq.~(4), and $\Sigma'(x)$ instead of $\Sigma''(x)$ in Eq.~(A2). The correct equations are
\begin{equation}
\tag{4 in \cite{KordyukPRB2005}}\label{ReS1}\Sigma'(\omega) = -\frac{v_F}{2 k_F} [k_m^2(\omega)-k_F^2] + \omega,
\end{equation}

\begin{equation}
\tag{A2 in \cite{KordyukPRB2005}}\label{KK2}\Sigma''(\omega) = -\frac{1}{\pi}\,\,PV\int_{-\infty}^\infty{\frac{\Sigma'(x)}{x - \omega}\,dx}.
\end{equation}

\end{document}